\documentclass[a4paper]{jpconf}
\usepackage{graphicx}
\usepackage{amsmath}
\usepackage{citesort}
\usepackage{color}
\newcommand{\sdo}{SrDy$_2$O$_4$}
\begin{document}
\title{Evidence for spin liquid ground state in \sdo\ frustrated magnet probed by $\mu$SR}

%\author{Nicolas Gauthier$^{1,*}$, Bobby Pr\'evost$^2$, Alex Amato$^3$, Chris Baines$^3$, Vladimir Pomjakushin$^4$, Andrea D. Bianchi$^2$, Robert J. Cava$^5$ and Michel Kenzelmann$^{1,\dagger}$ }
\author{N Gauthier$^{1,*}$, B Pr\'evost$^2$, A Amato$^3$, C Baines$^3$, V Pomjakushin$^4$, A~D Bianchi$^2$, R~J Cava$^5$ and M Kenzelmann$^{1,\dagger}$ 
}

\address{$^1$~Laboratory for Scientific Developments and Novel Materials, Paul Scherrer Institut, Villigen, Switzerland
$^2$~D\'epartement de Physique, Universit\'e de Montr\'eal, Montr\'eal, Canada
$^3$~Laboratory for Muon Spin Spectroscopy, Paul Scherrer Institut, Villigen, Switzerland
$^4$~Laboratory for Neutron Scattering and Imaging, Paul Scherrer Institut, Villigen, Switzerland
$^5$~Department of Chemistry, Princeton University, Princeton, New Jersey, USA}

\ead{$^{*}$~nicolas.gauthier@psi.ch, $^{\dagger}$~michel.kenzelmann@psi.ch}

\begin{abstract}
Muon spin relaxation ($\mu$SR) measurements were carried out on \sdo, a frustrated magnet featuring short range magnetic correlations at low temperatures. Zero-field muon spin depolarization measurements demonstrate that fast magnetic fluctuations are present from $T=300$~K down to 20~mK. The coexistence of short range magnetic correlations and fluctuations at $T=20$~mK indicates that \sdo\ features a spin liquid ground state. Large longitudinal fields affect weakly the muon spin depolarization, also suggesting the presence of fast fluctuations. For a longitudinal field of $\mu_0H=2$~T, a non-relaxing asymmetry contribution appears below $T=6$~K, indicating considerable slowing down of the magnetic fluctuations as field-induced magnetically-ordered phases are approached.
%. It is attributed to a fraction of the sample in which the moments appear static to the muons. 
%decreasing number of crystal-field levels are accessible by thermally-induced transitions
\end{abstract}

\section{Introduction}

In magnetically frustrated systems, competing interactions can inhibit long range order at low temperatures and enforce highly degenerate ground states. This can give rise to novel states of matter such as spin liquids, in which strongly fluctuating magnetic degrees of freedom persist to the lowest temperatures \cite{Balents2010}. For example, in the spin ices Dy$_2$Ti$_2$O$_7$ and Ho$_2$Ti$_2$O$_7$, well-known classical spin liquids, the low temperature dynamics are explained by the presence of magnetic monopoles \cite{Castelnovo2008,Jaubert2009,Fennell2009,Morris2009}. Other examples include Tb$_2$Ti$_2$O$_7$, where the magnetoelastic coupling is believed to play a major role in the spin liquid ground state \cite{Fennell2014a}, and MnSc$_2$S$_4$, where the spins fluctuate collectively between degenerate spiral states \cite{Bergman2007,Gao2016}. Muon spin relaxation ($\mu$SR) technique is a useful tool to study these types of systems since the experimental signatures of fluctuating moments, static disordered moments and static ordered moments are clearly distinguishable. It is therefore possible to distinguish a spin liquid from a spin glass or a long range ordered system.

The Sr\textit{Ln}$_2$O$_4$ (\textit{Ln}~=~Gd,~Tb,~Dy, Ho, Er, Tm and Yb) compounds form a family of frustrated rare-earth magnets \cite{Karunadasa2005,Li2014a}. They crystallize in an orthorhombic structure (space group \textit{Pnam}) with two inequivalent rare-earth ion sites, each forming one-dimensional zig-zag chains along the $c$-axis. In most of these compounds, the crystal electric fields generate different spin anisotropies at the two sites \cite{Fennell2014,Malkin2015} and lead to very anisotropic properties, as seen in magnetic susceptiblity measurements \cite{J.Hayes2012}. At very low temperatures, one dimensional correlations are observed in many members of this family \cite{Hayes2011,Young2013,Fennell2014,Wen2015} as a result of dominant magnetic interactions along the chains. The zig-zag chain is equivalent to the one-dimensional chain with nearest neighbour interaction $J_1$ and next-nearest neighbour interaction $J_2$, which for Ising spins is referred to as the axial next-nearest neighbour Ising (ANNNI) model \cite{Selke1988}. The model is frustrated for antiferromagnetic $J_2$ and this is the source of magnetic frustration in the Sr\textit{Ln}$_2$O$_4$ compounds. In \sdo, no evidence for a magnetic phase transition has been reported down to $T=50$~mK in zero magnetic field despite the emergence of broad magnetic scattering features below $T\approx4$~K in powder neutron diffraction \cite{Karunadasa2005}, reflecting the presence of a high degree of frustration. The specific heat features a broad maximum near $T=1.2$~K, indicative of magnetic correlations, but no sharp peak that would indicate a second order phase transition. At $T=50$~mK, powder neutron diffraction indicate that the system has quasi-one dimensional (1D) short range correlations~\cite{Fennell2014}. 

\section{Experimental details}
The $\mu$SR technique measures the time-dependent depolarization of spin-polarized muons that are implanted in a sample. The experimentally probed parameter is the asymmetry $A(t)$ given as:
\begin{equation}
A(t)=\frac{\alpha N_F(t)-N_B(t)}{\alpha N_F(t)+N_B(t)}
\end{equation}
where $N_F(t)$ and $N_B(t)$ are the number of detected positrons, arising from the weak-decay of the muon, in the forward (F) and backward (B) detectors respectively and $\alpha$ is a parameter taking into account the detector efficiency and geometry. The asymmetry $A(t)$ is proportional to the muon spin depolarization function $P(t)$ and $A(0)=a_0$ is the total initial asymmetry which is in part dependent on the instrument characteristics and typically $a_0\approx 0.25$. 

$\mu$SR spectra have been measured on the General Purpose Surface-Muon Instrument (GPS) and the Low Temperature Facility Instrument (LTF) at the Paul Scherrer Institut on powder samples of \sdo, prepared as reported in Ref. \cite{Karunadasa2005}. The spectra were collected in zero field from $T=5$~K to 300~K on GPS and from 20 mK to 9~K on LTF. Longitudinal field spectra have been measured on LTF up to $\mu_0H=2$~T. The $\alpha$ parameter was determined on GPS from a weak transverse field (TF) spectrum at $T=300$~K with $\mu_0H=0.005$~T, giving $\alpha=1.269(1)$. For LTF, it was determined from a weak TF spectrum at $T=20$~mK with $\mu_0H=0.010$~T, giving $\alpha=1.328(3)$. It is noteworthy that our measurements showed no temperature dependence of the TF spectra from $T=20$~mK up to 9~K on LTF. The obtained $\alpha$ parameter was used in the analysis of the zero-field (ZF) measurements. The value of this parameter changes in longitudinal field (LF) measurements due for instance to the trajectory curvature of the decay positrons in magnetic field. For the results presented here, the field dependence of $\alpha$ was evaluated by fitting it for the spectra at $T=9$~K in the paramagnetic regime from $\mu_0H=0$~to~2~T. The resulting values were used as fixed parameters for the other measured temperatures. 

\section{Experimental results}
\begin{figure}[!htb]
\centering
\includegraphics[scale=0.98]{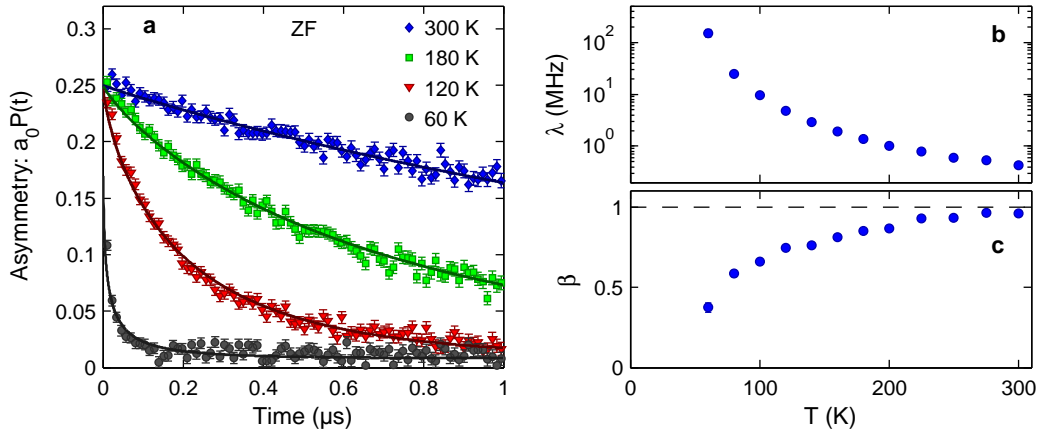}
\vspace{-0.3cm}
\caption{(a) $\mu$SR spectra of \sdo\ in zero field at various temperatures on GPS and the temperature dependence of the fitted parameters from Eq. \ref{asymmetry1}: (b) the relaxation rate $\lambda$ and (c) the exponent $\beta$.}
\label{musrZFGPS}
%\vspace{-0.2cm}
\end{figure}

At $T=300$~K in zero field, we observe that the initial asymmetry is close to 0.25 and decays exponentially with time (Fig. \ref{musrZFGPS}a). The relaxation rate increases with decreasing temperature. Below $T=60$~K the muon spin depolarization is so fast that most of the asymmetry relaxation occurs in the electronics dead time. This behaviour is observed down to $T=5$~K, the lowest reached temperature on GPS. A stretched exponential is the most appropriate function to describe the data over the full temperature range and the experimental spectra were fitted by:
\begin{equation}
A(t)=a_\text{s} e^{-\left( \lambda t \right)^\beta}+a_\text{b}
\label{asymmetry1}
\end{equation}
where $\lambda$ is the relaxation rate, $\beta$ is an exponent stretching the exponential function, $a_\text{s}$ is the sample initial asymmetry and $a_\text{b}$ is a constant asymmetry arising from muons missing the sample, with $a_\text{s}+a_\text{b}=a_0$. The spectra from $T=60$~K to 300~K were fitted simultaneously and the asymmetry values $a_\text{s}$ and $a_\text{b}$ were taken as temperature-independent parameters. The fits show that the relaxation rate $\lambda$ increases with decreasing temperature (Fig. \ref{musrZFGPS}b) while the exponent $\beta$ is almost one at $T=300$~K and decreases with decreasing temperature (Fig. \ref{musrZFGPS}c). 

For $\beta=1$, the relaxation corresponds to a simple exponential where the relaxation rate can be defined in the fast fluctuations limit as $\lambda=2 \Delta^2/\nu$ \cite{Reotier1997}. The parameter $\Delta$ is proportional to $\sqrt{\left< B_\text{loc}^2 \right>}$, where $B_\text{loc}$ is the dynamic local internal field at the muon implantation site, and it thus is a measure of the field distribution width. The parameter $\nu$ is the fluctuation rate of $B_\text{loc}$. In systems with large magnetic moments, $\Delta$ can be expected to be large and responsible for the fast depolarization of the muon spin. In the paramagnetic regime, $\Delta$ should be weakly temperature dependent while $\nu$ should reduce at lower temperature, as a decreasing number of crystal-field levels are accessible by thermally-induced transitions. This implies an increasing $\lambda$ with decreasing temperature, in agreement with our observation. 

A value of $\beta$ different from unity can be interpreted has the presence of a distribution of relaxation rates $\lambda$ \cite{Campbell1994}. At $T=300$~K the exponent $\beta$ is close to one and this system is well described by a single relaxation rate. As the temperature decreases, $\beta$ decreases, suggesting a broadening of the relaxation rate distribution, from multiple fluctuation rates $\nu$ and/or multiple field distribution widths $\Delta$. The presence of two inequivalent magnetic sites with different single ion anisotropy could explain such an effect. 

\begin{figure}[!htb]
\centering
\includegraphics[scale=0.98]{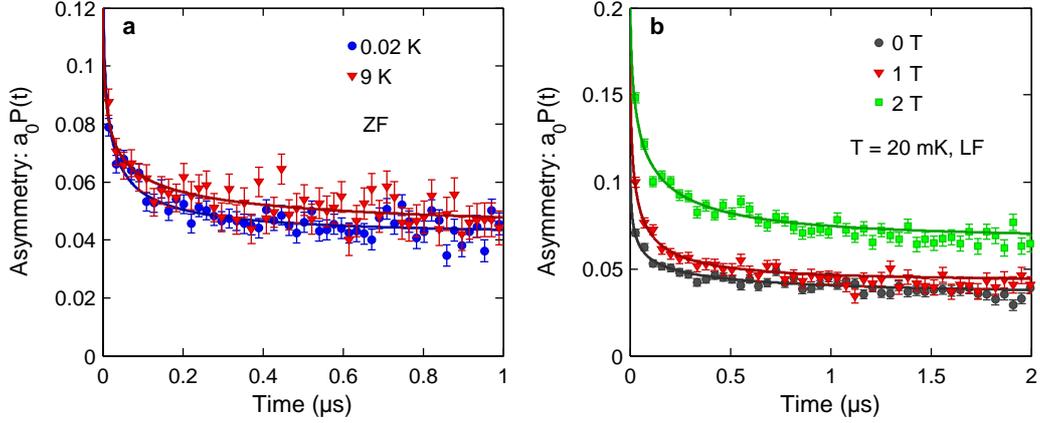}
\vspace{-0.3cm}
\caption{$\mu$SR spectra of \sdo\ (a) in zero field at $T=20$~mK and 9~K on LTF and (b) in longitudinal fields at $T=20$~mK on LTF.}
\label{musrZFLTF}
%\vspace{-0.2cm}
\end{figure}

$\mu$SR spectra were also measured at LTF from $T=9$~K down to 20~mK and no significant temperature dependence was observed (Fig. \ref{musrZFLTF}a). The temperature range overlaps with the GPS measurements and suggest that the muon spin polarization on the LTF spectra is too fast to be reliably measured. This indicates that fluctuations remain down to 20~mK, in agreement with the absence of long range order in this compound \cite{Fennell2014}.

% A recovery of $\frac{1}{3}$ of the initial asymmetry, which is not observed, would be expected in the presence of static moments. \textcolor{red}{do we really want to say this? even if there were a 1/3 of the initial asymmetry, it could also decay rather quickly through some transverse fluctuations, I think we don't gain much by claiming this}

%The background value $a_\text{b}$ was fixed to 0.0418, determined from the amplitude of the oscillations in the weak TF spectrum at 0.02~K. The remaining asymmetry is attributed to the tail of the stretched exponential. Therefore, $a_\text{s}$ was fixed to $a_\text{s}=a_0-a_\text{b}$, where $a_0=0.25$. The fitted parameters $\lambda$ and $\beta$ have a weak temperature dependence (Fig. \ref{musrZFLTF}b,c). This result must be taken with care since is not very reliable. 

Longitudinal field measurements give more insights into the low temperature spin dynamics. Typically for static internal fields, the application of a sufficiently strong longitudinal field will dominate the muon spin depolarization and effectively decouple it from the sample internal fields. For dynamic internal fields, the application of a longitudinal field usually has a smaller effect. In \sdo, the muon spin depolarization is weakly affected by a longitudinal field of $\mu_0H=1$~T while a non-relaxing asymmetry contribution attributed to static moments emerges at $\mu_0H=2$~T (Fig.~\ref{musrZFLTF}b). This is strong evidence of the presence of spin dynamics for $\mu_0H\leq1$~T. 
%The similarity of the spectrum for $H=1$~T at $T=20$~mK with the one at 9~K in the paramagnetic regime (Fig. \ref{musrLF1}a) is a further evidence that magnetic fluctuations are present down to 20~mK.

\begin{figure}[!htb]
\centering
\includegraphics[scale=0.98]{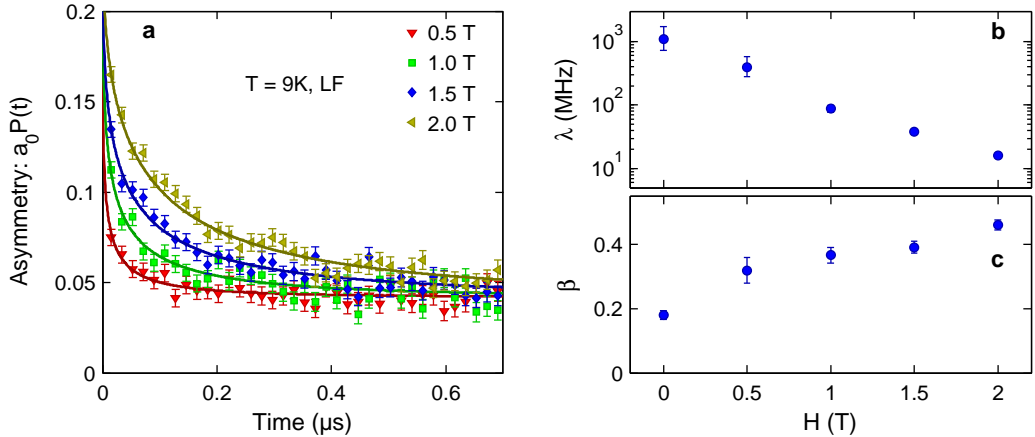}
\vspace{-0.3cm}
\caption{(a) $\mu$SR spectra of \sdo\ at $T=9$~K in longitudinal fields on LTF and the magnetic field dependence of the fitted parameters from Eq. \ref{asymmetry1}: (b) the relaxation rate $\lambda$ and (c) the exponent $\beta$.}
\label{musrLF1}
%\vspace{-0.2cm}
\end{figure}

Longitudinal-field measurements at $T=9$~K for fields up to $\mu_0H=2$~T, shown in Fig.~\ref{musrLF1}a, show the presence of spin dynamics, as expected in the paramagnetic regime. The initial asymmetry increases with increasing field, indicating that the relaxation rate is reduced with field. The spectra were fitted with Eq. \ref{asymmetry1} by fixing $a_0=0.25$, the expected total initial asymmetry, and $a_\text{b}=0.0418$, as determined from the amplitude of the oscillations in the weak TF spectrum at $T=0.02$~K in zero field. The fitted parameters $\lambda$ and $\beta$ evolve smoothly with field (Fig. \ref{musrLF1}b,c). The increasing value of the exponent $\beta$ suggests a narrowing of the relaxation rate distribution. The decreasing value of the relaxation parameter $\lambda$ indicates a gradual decoupling of the muon spin polarization from the sample, while the polarization is increasingly affected by the external field. The effect observed here is relatively weak in comparison to the strength of the applied field, indicating that the magnetism at $T=9$~K is dynamic up to $\mu_0H=2$~T.

\begin{figure}[!htb]
\centering
\includegraphics[scale=0.98]{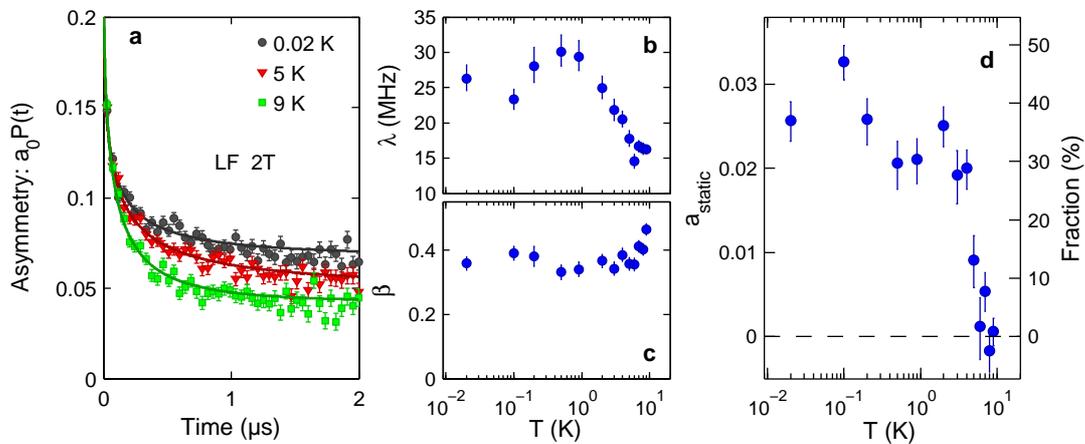}
\vspace{-0.3cm}
\caption{(a) $\mu$SR spectra of \sdo\ at various temperatures in a longitudinal field of $\mu_0H=2$~T on LTF and the temperature dependence of the fitted parameters from Eq. \ref{asymmetry2}: (b) the relaxation rate $\lambda$, (c) the exponent $\beta$ and (d) the time-independent constant $a_\text{static}$ (left scale). The corresponding sample fraction which has static moments is also represented on the right scale.}
\label{musrLF2}
%\vspace{-0.2cm}
\end{figure}

A temperature dependence of LF spectra at $\mu_0H=2$~T shows the emergence of the non-relaxing asymmetry contribution below $T=6$~K (Fig. \ref{musrLF2}). For static fields the muon polarization at long times should lead in a powder sample to a constant $\frac{1}{3}$ of the initial asymmetry. The observed non-relaxing asymmetry is therefore attributed to a sample fraction where the moments appear static to the muons, i.e. the fluctuation rate is too slow to be felt by the muons which have a lifetime of 2.2~$\mu$s. Therefore, the Eq. \ref{asymmetry1} was modified to take into account this new term:
\begin{equation}
A(t)=a_\text{s} e^{-\left( \lambda t \right)^\beta}+a_\text{static}+a_\text{b}
\label{asymmetry2}
\end{equation}
where $a_\text{static}$ is a non-relaxing constant accounting for the static fraction of the sample, and the total asymmetry is fixed such that $a_\text{s}+a_\text{static}+a_\text{b}=a_0$. The value $a_\text{b}$ was taken from the zero field spectrum and kept fixed for all temperatures. The temperature dependence of the stretched exponential parameters are shown in Fig. \ref{musrLF2}b,c. The exponent $\beta$ is fairly constant but the relaxation rate $\lambda$ increases below $T=6$~K. Under that same temperature, the constant $a_\text{static}$ becomes non-zero and increases with decreasing temperature, reaching a value of about one tenth of the initial polarization (Fig.~\ref{musrLF2}d).

\section{Discussion and summary}

The high temperature $\mu$SR spectra of \sdo\ share many similarities with the spin ice Dy$_2$Ti$_2$O$_7$. In both case, a stretched exponential describe the data well and the fitted parameters have a similar tendency with decreasing temperature: (i) an increase of the relaxation rate $\lambda$ and (ii) a reduction of $\beta$ with a maximum of unity at room temperature \cite{Lago2007}. Below $T\approx60$~K, the muon spin depolarization is so fast that it can not be measured in both experiments, probably as a consequence of a broad internal field distribution (large $\Delta$) due to the large magnetic moment of Dy$^{3+}$ of the order of 8-10~$\mu_B$ \cite{Fennell2014}. 

At low temperatures, \sdo\ and Dy$_2$Ti$_2$O$_7$ show a very different muon spin relaxation. While all the muon spin is mostly depolarized below $T\approx60$~K, Dy$_2$Ti$_2$O$_7$ features a recovery of about one third of the spin polarization below $T\approx40$~K. This was explained by the dominance of slow magnetic fluctuations that appear static in the time window of the muon decay \cite{Lago2007}. Firstly, this leads to a large field distribution at the muon sites and the associated fast relaxation of two thirds of the muon polarization. Secondly, this also leads to a pinning of the muon spin parallel to the magnetic field at the muon site, recovering one third of the muon polarization. This recovered muon polarization decays on a much longer time scale with a relaxation rate directly proportional to magnetic fluctuations frequency.

In contrast, the muon spin polarization in \sdo\ features a fast relaxation in excess of 150~MHz at all temperatures below $T=60$~K, showing the presence of fast magnetic fluctuations. The origin of these magnetic fluctuations may be thermal excitations to the low-lying crystal-field levels at somewhat intermediate temperatures, and short-range magnetic correlations at lower temperatures, also observed by neutron powder diffraction \cite{Fennell2014}. The longitudinal field measurements also confirms that fluctuations are present for $\mu_0H\leq1$~T at $T=20$~mK, where thermal fluctuations are not expected to play any role.

%\textcolor{red}{Assuming that the magnetic fluctuations slow down as in Dy$_2$Ti$_2$O$_7$, the expected slow decay of one third of the muon polarization is still too fast to be observed in \sdo\ and the magnetic fluctuations frequency can be estimated to be higher than 150~MHz down to $T=20$~mK. }

%the signal is observed under 40~K corresponding to $\frac{1}{3}a_0$. The $\frac{2}{3}a_0$ transverse contribution is still too fast to be measured due to a large $\Delta$. The $\frac{1}{3}a_0$ longitudinal contribution is recovered due to the presence of very slow fluctuations. The relaxation of this contribution is described in the slow fluctuations limit with $P(t)=\frac{1}{3}e^{-\frac{2}{3}\nu t}$ where $\nu$ is the fluctuation rate. Such a behaviour is not observed in \sdo\ and based on this, we estimate that the fluctuation rate is of the order of 200~MHz or higher. %While the depolarization observed at low temperatures can be accounted for by the tail of a stretched exponential, it is likely that the nuclear spin system also contributes to this depolarization. Therefore, the fitted parameters do not allow any further claim on the physics of the system than the presence of a very fast depolarization down to 20~mK, due to fluctuations of the moments. 
%The coexistence of these fluctuations and magnetic correlations, as evidenced , indicates that \sdo\ is to be considered as a spin liquid. 

For a longitudinal field of $\mu_0H=2$~T, the non-relaxing polarization emerging below $T=6$~K reaches about one tenth of the initial polarization at $T=20$~mK, corresponding to a sample fraction of $\sim35\%$ with static moments. These static moments could be ordered or disordered but field-induced phases evidenced by magnetization plateaus \cite{J.Hayes2012} suggest the former. The fact that only part of the sample appears static is probably the result of the highly anisotropic properties of \sdo~\cite{J.Hayes2012}.

In summary, we present the results of a $\mu$SR study on \sdo\, indicating the presence of fluctuations down to the $T=20$~mK and the emergence of a static fraction in longitudinal field of $\mu_0H=2$~T. Due to the large moments of Dy$^{3+}$, broad field distributions are expected at the muon sites leading to a very fast muon spin depolarization. In the present study, 
this depolarization becomes too fast to be measured accurately below $T=60~$K in zero field.
The absence of a slowly-relaxing muon spin component on the order of one-third of the initial polarization at low temperatures is evidence that fast fluctuations remain the dominant mechanism for the muon spin relaxation in \sdo, suggesting a spin liquid ground state. %The non-relaxing fraction of the spectrum with a longitudinal field of $H=2$~T below $T=6$~K indicate that static local fields develop in a fraction of the sample.

\ack
The authors acknowledge useful discussions with Hubertus Luetkens and Daniel G. Mazzone. This research received support from the Natural Sciences and Engineering Research Council of Canada (Canada).

\section*{References}

\bibliographystyle{iopart-num}
\bibliography{biblioV2}

\end{document}